\documentstyle[11pt,mrs2001,epsfig]{article}

\begin{document}

\title{Voids in Galaxy Redshift Surveys}

\author{F. Hoyle $^1$, M. S. Vogeley$^2$}
\affil{Drexel University, 3141 Chestnut Street, Philadelphia, PA, 19104, USA}
\affil{$^1$hoyle@venus.physics.drexel.edu, $^2$vogeley@drexel.edu}

\begin{abstract}
One of the most striking features in galaxy redshift surveys
is the ubiquitous presence of voids. However,
voids have not been extensively studied due to observational
limitations. Until recently, galaxy redshift surveys included only 
a few voids of diameter $> 30 h^{-1}$Mpc.

The first step in studying voids is the identification
process.  We outline a method for detecting voids
(based on the method of El-Ad and Piran 1997, EP97). We apply
it to the PSCz survey \cite{saund}, and the Updated
Zwicky Catalog \cite{falco}. We find that voids have typical
diameters of $\sim 30 h^{-1}$Mpc and are very underdense regions
with $\delta \rho/\rho \sim -0.95$. Up to 40\% of volume of the
Universe is occupied by voids.
We discuss the results from these surveys and our detection
algorithm's usefulness for future surveys.

The next generation of surveys (such as the Sloan Digital
Sky Survey and 2dFGRS) will improve this situation. The surveys will extend
to greater depths, allowing a more statistically complete
sample of voids to be obtained and the accompanying
digital imaging will provide accurate photometry of
fainter objects. This will allow us to gain greater
insight into the nature of voids, search for possible void
galaxies and maybe place constraints on cosmological parameters
as voids play a critical role in the evolution of large scale
structure.

\end{abstract}

\section{Introduction}

Our knowledge of the distribution of luminous matter in the Universe has increased dramatically over the last two decades. Both angular catalogues and redshift surveys now contain many thousands of galaxies and have allowed us to study their clustering properties in great detail. However, one feature of redshift surveys has been somewhat overlooked. A cursory glance at the distribution of galaxies in redshift surveys or angular catalogues reveals that there are vast parts of any survey that seem to be avoided by galaxies. These are termed voids. They are a major feature in the large scale distribution of galaxies in all wide angle, densely sampled redshift surveys. 

Until recently, voids have not been studied in great detail. This was mostly due to limitations on the sizes of the surveys. The $\sim$ 30 $h^{-1}$Mpc \footnote{we adopt the convention that $H_{\circ}=100 h $km s$^{-1}$ Mpc$^{-1}$} diameter of voids is a significant fraction of the depth of many redshift surveys of the local universe \cite{elpdc96} \cite{elpdc97} \cite{elp97}. A search for voids has so far been made on the Southern Sky Redshift Survey \cite{elpdc96}, the IRAS 1.2 Jy Survey \cite{elpdc97} and the Las Campanas Survey \cite{muller}. Approximately twelve voids have been detected in these nearby surveys. 

Observed voids contain few galaxies regardless of the detection criteria. However, cold dark matter models predict that there should be matter, and hence maybe galaxies, within voids (see Benson et al. and Somerville et al. in this volume for a discussion of the substructure problem). Yet, studies of different types of galaxies show that they all seem to trace the same structures. Either we are missing a population of galaxies that reside in the voids or there is a problem with the models producing too much mass on small scales \cite{Peebles}.

A first step in quantifying the properties of voids and void galaxies is to compile a catalogue of detected voids from as many redshift surveys as possible. Here we adapt the method of EP97 in order to search for voids in the PSCz Survey \cite{saund} and the Updated Zwicky Catalog \cite{falco}. We describe the surveys in Sections \ref{sec:surv}, the algorithm in Section \ref{sec:algo} and present our results in Section \ref{sec:res}.

\section{The Surveys}
\label{sec:surv}

The PSCz Survey \cite{saund} is a redshift survey of galaxies that is based on the catalogue of detections with the Infra-Red Astronomical Satellite. It covers the whole sky apart from areas with incomplete IRAS data and high optical extinction, such as the plane of the Galaxy. The PSCz survey covers 84\% of the sky and contains a total of 15,411 galaxies, with a median redshift of 0.028 ($82.2 h^{-1}$Mpc). 

The Updated Zwicky Catalog (UZC) \cite{falco} includes results of the Center for Astrophysics Redshift Survey \cite{Huchra} \cite{geller} and contains a total of 18,633 galaxies with redshifts. The survey covers two main survey regions; $ \alpha > 20^h, \alpha < 4^h$ and $8^h > \alpha < 17^h$ both with $-2.5^{\circ} < \delta < 50^{\circ}$. The volume limited sample which contains the most galaxies extends to $z_{\rm max}$=0.025 ($73.6 h^{-1}$Mpc) and contains 3518 galaxies.

We consider different samples from the two surveys to check the robustness of results, in particular, we check what effect the wall/field criteria has on the detection of voids and how much the volume of voids is underestimated by insisting that voids lie completely within the survey geometry. We also compare the two surveys to see if the same voids are detected and if the properties of voids in overlap regions are similar even though one survey is IR selected, the other optical. 

\section{The Void Finding Algorithm}
\label{sec:algo}

Here we give a brief overview of the steps involved in our void finding algorithm. Each of the steps are discussed in detail in the subsequent sections. 
The void finding algorithm we adopt is very similar to the method used by El-Ad \& Piran ({\tt Void Finder}) \cite{elp97}. The main features of the algorithm are:
\begin{itemize} 
\item it is based on the point distribution of galaxies;
\item allows for so called `field galaxies' - voids may not be completely empty;
\item voids have to be larger than a minimum threshold
\end{itemize}

\subsection{Wall and Field Galaxies}

The first stage in identifying voids is to determine which galaxies are classified as field galaxies and which are wall galaxies. Here we follow the method of EP97. First we calculate the mean distance, {\it d}, to the $n^{\rm th}$ galaxy and the standard deviation on this value, $\sigma$. We then specify a length {\it l$_n$} such that any galaxy that does not have $n$ neighbours within a sphere of radius {\it l$_n$} is classified as a field galaxy. The value of {\it l$_n$} we adopt is given by {\it l$_n$} = {\it d} + 1.5$\sigma$ and we set $n$=3. For our main samples, the values of {\it d} and {\it l$_3$} that we obtain for the PSCz survey are 3.42$h^{-1}$Mpc and 7.19$h^{-1}$Mpc respectively while for the UZC we obtain 3.46$h^{-1}$ and 6.52$h^{-1}$Mpc. A choice of {\it n}=3 means that $\sim$ 10\% of galaxies are classified as field galaxies for both surveys. We compare with the results for a sample where we do not make the wall/field split to see what effect the choice of $n$ makes since the fraction of field galaxies depends critically on this choice. We count the number of field galaxies that lie in the voids to calculate the density of voids.

\subsection{Finding the Voids}
\subsubsection{Identifying Holes}

\begin{figure} 
\plottwo{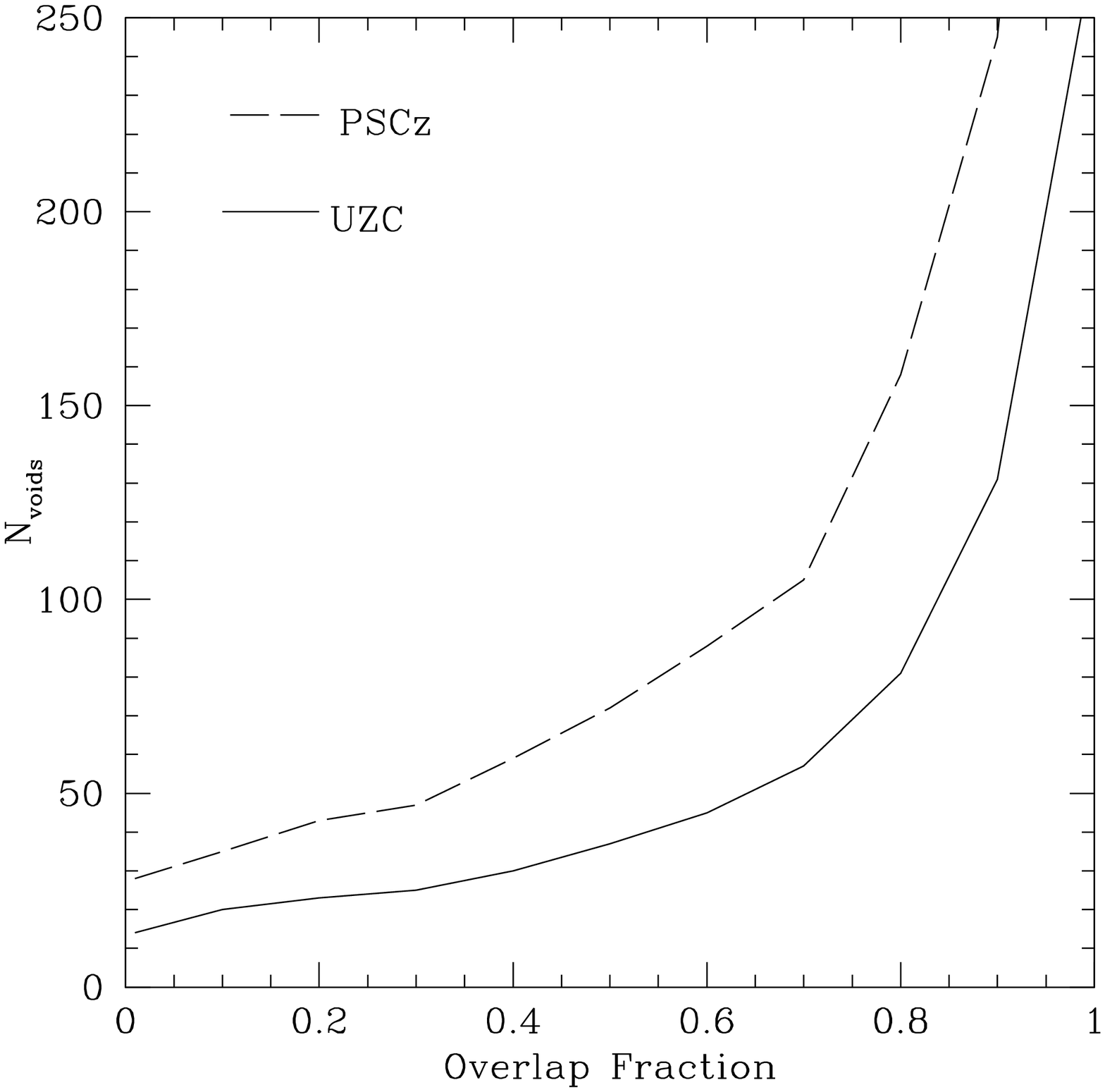}{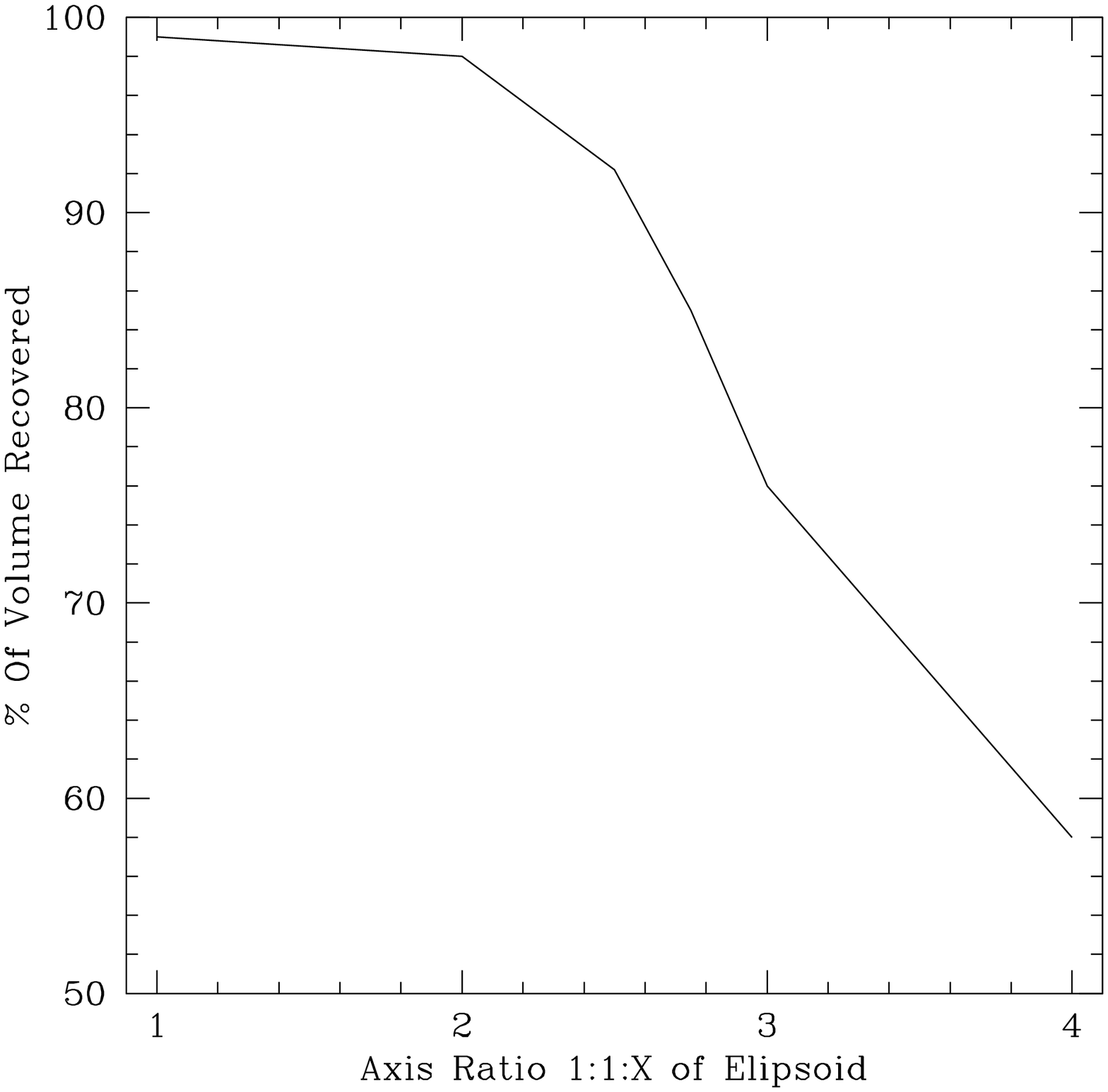}
\caption{The left hand plot shows the number of voids found in each survey when voids that overlap by the fraction shown are classified as distinct voids. The right hand plot shows how well we are able to recover the volume of voids with axis ration 1:1:X.}
\label{fig:fig1}
\end{figure}

The way we detect the voids is similar to the {\tt void finder} algorithm of EP97. First we place the wall galaxies onto a three dimensional grid. The fineness of the grid defines the minimum size void that can be detected. If each grid cell has length l$_{\rm cell}$, then all voids with radius r = $\sqrt{3}$l$_{\rm cell}$ will be found. 

Each empty cell (we will refer to these as holes) is considered a possible void. Our method finds the maximal sphere than can be drawn in the hole, starting from the first empty cell. We place a sphere at the centre of the cell and grow its radius until we find a galaxy that is just on the edge of it. We then find the vector that connects this galaxy with the centre of the hole and move the centre of the spherical hole along this direction, away from the first galaxy, growing the radius to keep the first galaxy on its surface, until a second galaxy is also on the surface. We next find the vector that bisects the line joining the two galaxies and move the hole in this direction until a third galaxy is found, as before. The final step is to grow the hole out of the plane formed by the first three galaxies. At this stage we keep track of all the holes with radii larger than the value of the search radius used to classify field and wall galaxies, $l_{3}$.

\subsubsection{Determining a Void}

Finding the holes is a robust process. Deciding which holes form unique voids requires more care. Our definition of a void is slightly different from that of EP97. First we sort the holes by radius, the largest first. The largest hole found is automatically a void. We test the second hole. If it overlaps the first by more than F\% in volume, then we say it is a member of the first void rather than a new void. If not then it forms a separate void. We then check the third hole. If it does not overlap a previous void by F\%, it is a distinct void. If it overlaps only one previously detected void by F\%, it becomes a member of that void. If it overlaps more than one previous void by F\%, it is rejected as it is linking together two larger voids. We continue like this for all holes with radii larger than 10$h^{-1}$Mpc. 

We have investigated how much holes should overlap and yet be considered separate voids. In Figure 1a we show the number of voids we find for the PSCz (dashed line) and the UZC (solid line) as a function of the overlap fraction for which the hole is still considered a separate void. The number of voids converges to within a factor of two if the overlap fraction is $<$30\%. We fix the fraction to be 10\%. Therefore, if a sphere with radii larger than 10$h^{-1}$Mpc overlaps a larger void by more than 10\% of its volume, we merge the sphere into the void. If the overlap fraction is less than 10\% of the smaller voids volume, we deem the sphere to be a distinct void. Visually, using our criteria of 10\% overlap, voids appear to be distinct regions of the survey. 

We set a threshold of 10$h^{-1}$Mpc for the minimum size of voids, this value being chosen for two reasons. It is larger than the search radius for defining field galaxies, $l_3$. This helps ensure that we don't identify gaps in the walls as voids. It is also the value at which the significance of detecting voids in the both the PSCz and UZC drops below 80\%, as discussed later.

\subsubsection{Enhancing the Volume}

We next enhance the volume of the void. We consider the holes that have radii less than the threshold of 10$h^{-1}$Mpc but greater than the radius for the wall/field galaxy criteria (in the case where we do not make the wall/field galaxy cut we use the value we would have used to make the cut as the minimum sphere radius). Any of these holes that overlap the maximal void sphere by 50\% of the smaller holes volume are also considered members of the void. If the hole overlaps with more than one void by 50\% then it is not added to either of the voids as this would link two voids together that we wish to keep seperate. If the hole is isolated it cannot be classified as a seperate void as it is smaller than the threshold we use for void classification. We have varied the overlap fraction and we find that the volume of voids are robust to within $\sim$50\% if the minimum overlap fraction varies from 0.3-0.8.

We are likely to underestimate the volumes of voids as we only consider holes that have radius greater than the search radius, $l_3$. If voids are highly elliptical then we will not detect the volume at the `corners' of the ellipse. We test this by generating data containing mock voids of known elliptical shape. We generate ellipsoids with volume 15,000$h^{-3}$Mpc$^3$ that have axis ratios 1:1:X (i.e. if X is 1 then the void is spherical with radius 15.3$h^{-1}$Mpc). We then run the simulated data through our void finding algorithm and compare the volume obtained with the known volume of the void. The results are shown in Figure 1b. If voids are spherical in shape then we recover 100\% of its volume. The more elongated it becomes, the less of the volume we detect. This becomes significant if one of the axes is elongated by more than a factor of 2.5 than the others.

\subsection{Significance of the Voids}

Again we use the method of EPC97 to asses the significance of voids. The confidence level with which we detect a void is given by $p = 1 - N_{\rm Poisson}/N_{\rm Survey}$
where $N_{\rm Poisson}$ and $N_{\rm Survey}$ are the number of voids found in the Poisson realisations and the survey sample under consideration.
The closer $p$ is to 1, the less likely a void could occur in a random distribution.

\begin{figure} 
\plottwo{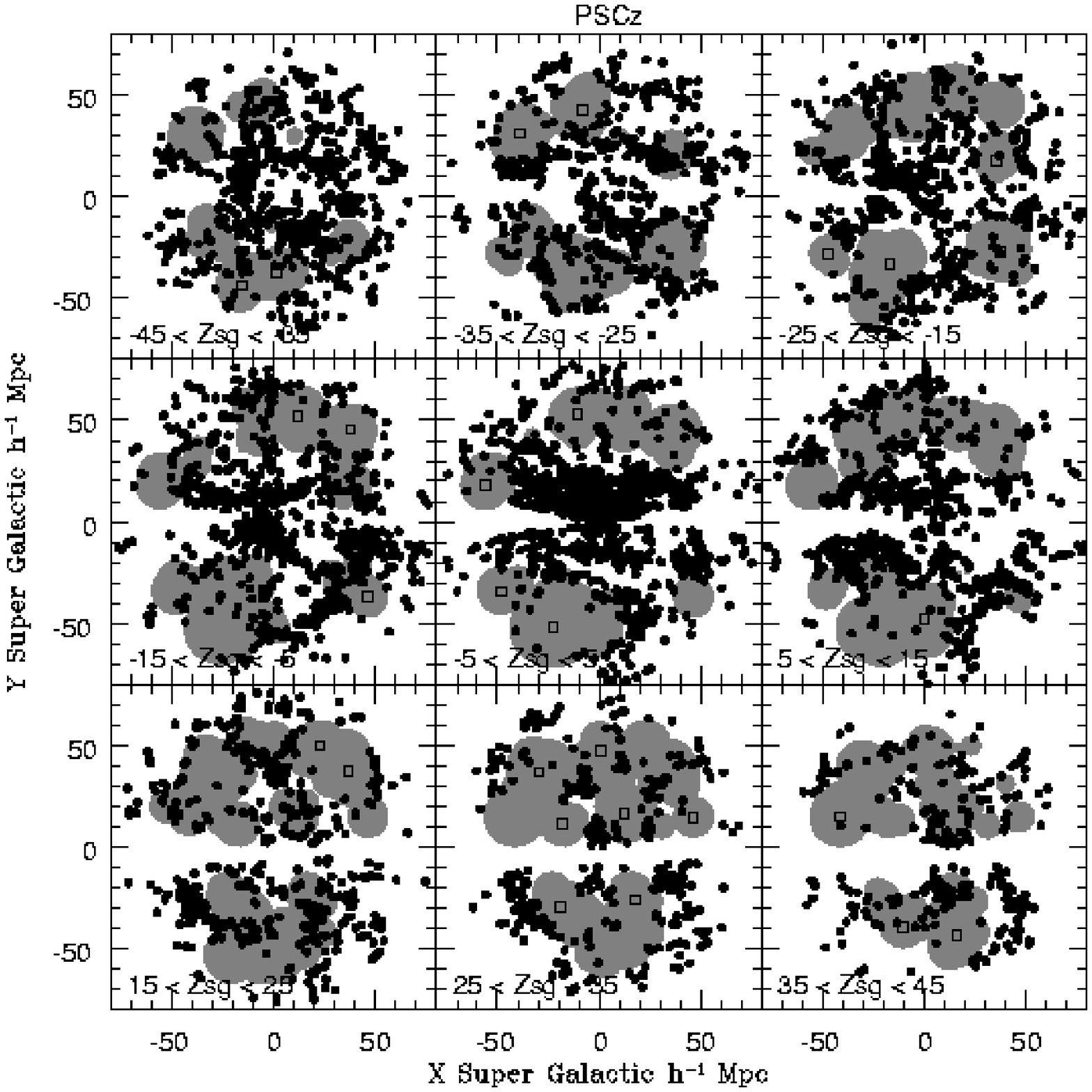}{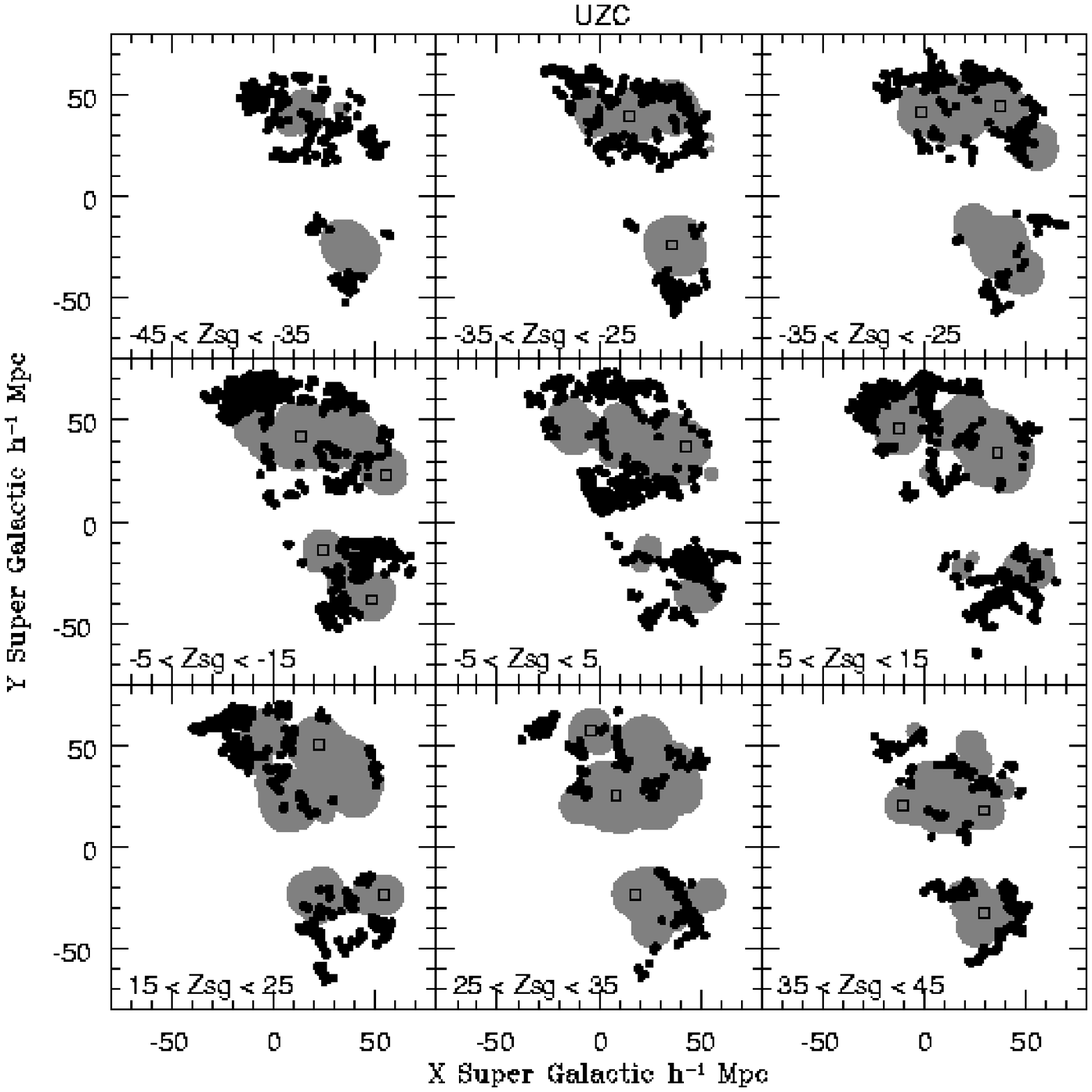}
\caption{We show the supergalactic coordinates (x,y) for different values of z. Each panel shows a 10$h^{-1}$Mpc slice. The shaded regions are the voids. The points are the wall galaxies and the empty squares show the void centres. No wall galaxies are found in a void.}
\label{fig:cfavoids}
\end{figure}

\section{Results}
\label{sec:res}

Voids in the main PSCz and UZC sample are shown in Figure 2. In both cases, the points represent the wall galaxies and the shadings represent areas that are covered by voids and the open squares show the void centres. 

We find 35 voids in the PSCz survey and 20 in the UZC with r$ > 10 h^{-1}$Mpc. We construct mock PSCz and UZC samples and find that we detect an average of 7.2 voids with r$ > 10 h^{-1}$Mpc in Poisson realisations of the PSCz sample and 3.9 voids with r$ > 10 h^{-1}$Mpc in Poisson realisations of the UZC. This implies that the voids in both the PSCz and UZC are significant at the 80\% confidence level. 

The range of sizes of the voids between the two surveys is comparable. The largest maximal sphere in the PSCZ survey has radius 17.8$h^{-1}$Mpc, whereas in the UZC survey the largest hole has radius 14.7$h^{-1}$Mpc. The average void in the PSCz survey has a volume of 15200$h^{-3}$Mpc$^3$ whereas the average void in the UZC survey has a slightly smaller volume of 12800$h^{-3}$Mpc$^3$. Using our criteria for identifying voids, the total volume fraction occupied by voids in the PSCz and UZC surveys is 30\% and 40\% respectively.

In the case where we differentiate between wall and field galaxies, we find that voids have a typical overdensity of $\delta \rho / \rho$ = -0.94 in the case of the PSCz survey and -0.98 in the case of the CfA survey. These values are very low, even with 10\% of galaxies classified as field galaxies, probably because the field galaxies lie close to the structures traced by the wall galaxies and are therefore not detected within the void volume. If we do not apply the field/wall criteria, we find 90\% of the same voids. We miss one or two of the voids if we don't define some galaxies as field galaxies because the few field galaxies in the low density environments restrict the size of the maximal sphere.

We compare the volumes of the voids found in our main sample with voids found in samples that extend 20$h^{-1}$Mpc in depth beyond the depth of our main samples. This test shows that we underestimate the volume of voids by a larger factor the further out the void is found but the volumes are found to agree within a factor of $\sim 2$ for most of the voids. 

Full details of the method and detected voids are given in Hoyle \& Vogeley.

\section{Conclusions}
\label{sec:conc}

We have demonstrated that our technique gives robust results in the sense that different samples from the same survey yield the same voids and we detect the same voids in different redshift surveys. As an extension to the work of EP97 and EPC97 we have quantified the effect of our void definition on the number of voids detected within a survey and we have provided estimates of how accurately we are able to recover the volume of voids. Using criteria that detect voids with typical underdensity $\delta \rho/\rho < -0.9$, we find that up to 40\% of the volume in the surveys under consideration is found in void regions. This is consistent with the findings of EPC97 and shows that voids are indeed a large part of the Universe.

The next generation of surveys, the 2dF Galaxy Redshift Survey \cite{peacock} and, in particular, the Sloan Digital Sky Survey \cite{york} will aid our understanding of voids. Both of these surveys cover a larger area than the UZC, although not quite as large an area as the IR surveys. The SDSS will cover a quarter of the sky in one contiguous area which will be especially useful for void detection. Both the 2dFGRS and the SDSS will reach fainter magnitude limits than previous surveys, approximately 4 magnitudes deeper than the UZC. This will allow us to construct volume limited samples with more galaxies, which extend to greater depths. Also, the SDSS bright red galaxy sample will be approximately volume limited out to $z_{\rm max} \sim 0.5$ which increases the depth by more than an order of magnitude, although due to the sparsity of galaxies in the BRG sample the voids will not be detected with as high a confidence as they will in the main galaxy sample of the SDSS. 

\vspace{0.2cm}
We acknowledge the support of NSF grant AST-0072101 and a grant from the John Templeton Foundation

\vfill
\end{document}